# Towards Artefact-based Requirements Engineering for Data-Centric Systems

Tatiana Chuprina, Daniel Mendez[a,b] and Krzysztof Wnuk[a]

[a]*Blekinge Institute of Technology, Karlskrona, Sweden*
[b]*fortiss GmbH, Munich, Germany*

**Abstract**
Many modern software-intensive systems employ artificial intelligence / machine-learning (AI/ML) components and are, thus, inherently data-centric. The behaviour of such systems depends on typically large amounts of data processed at run-time rendering such non-deterministic systems as complex. This complexity growth affects our understanding on needs and practices in Requirements Engineering (RE). There is, however, still little guidance on how to handle requirements for such systems effectively: What are, for example, typical quality requirements classes? What modelling concepts do we rely on or which levels of abstraction do we need to consider? In fact, how to integrate such concepts into approaches for a more traditional RE still needs profound investigations. In this research preview paper, we report on ongoing efforts to establish an artefact-based RE approach for the development of data-centric systems (DCSs). To this end, we sketch a DCS development process with the newly proposed requirements categories and data-centric artefacts and briefly report on an ongoing investigation of current RE challenges in industry developing data-centric systems.

## 1. Introduction

Requirements Engineering (RE) for software-intensive system faces already many challenges [1]. When considering the development of AI-/ML-intensive software systems, where run-time behaviour is largely non-deterministic, the complexity growth renders RE even more difficult to handle. In fact, we still lack a clear understanding about how to efficiently consider such technologies in a system development life cycle and, thus, also in RE.

AI/ML technologies are data-centric by nature and such *Data-centric Systems (DCSs)* process a great amount of data at run-time. This peculiarity of DCSs requires new RE approaches. For example, a key asset for a DCS is a training data set which largely dictates the system behaviour; for instance, a self-driving car should learn how to identify static and dynamic obstacles such as traffic lights or pedestrians on a street. New applied technologies (e.g., neural networks) also bring additional quality attributes such as explainability or fairness[2]. This further renders requirements specifications and validations challenging.

The classification of requirements and their underlying modelling concepts is already in strong contrast to what is typically addressed under the umbrella of traditional RE approaches. Overall, we have still little knowledge about how to efficiently integrate data-centric aspects into RE. Our long-term objective is to tackle this gap by establishing a conceptual (reference) model of relevant artefacts, contents, and relationships to guide RE for DCSs. To this end, however, we first need a better understanding of the current industrial situation in the DCS development including how practitioners handle requirements for DCSs, what processes they apply to this end, and what problems they encounter. This allows for the problem-driven research we envision.

In section 3, we briefly highlight challenges we distilled so far from an ongoing investigation. Moreover, in section 4, we will give first insights into the artefact-based RE reference model we are working on. That model development is steered by our positive experiences made with artefact orientation [3]. Both shall outline our research roadmap which, hopefully, serves as a basis to foster important discussions in the direction of an effective RE for DCSs.

## 2. Related Work

The body of knowledge in software engineering for AI-intensive software systems is growing fast with many promising research directions. Here, we focus on an overview of work we consider directly related to our research.

Amershi [4], for instance, report on experiences at Microsoft when integrating ML technologies into a development workflow. Their insights corroborate the complexity growth of the development process when involving ML components with respect to, inter alia, version control and data management or ML models reuse. The authors conclude that data aspects are cutting through the whole development process and, in consequence, contribute to the difficulties separating ML-intensive components from each other and from software-intensive components. Another lesson learned in software engineering with focus on the integration of ML-component is presented by Arpteg [5]. The focus of their study was however on details regarding the application of deep learning (DL) techniques by different companies of varying sizes. The authors further elaborated on challenges faced during development, but also during production and at organisational level.

Another study we consider relevant is one conducted by Ishikawa [2]. They present a quantitative view on challenges often encountered during engineering of ML-intensive systems derived from a questionnaire-based survey with 278 participants. This study identifies the lack of appropriate approaches and tools for the engineering of these systems.

Lwakatare report on the results of a case study conducted with six companies in different domains, such as automotive and telecommunication [6]. The results of the study uncover difficulties in software engineering for ML which relate to inefficient techniques for training, validation, and deployment of ML models as well as challenges with the proper data set collection. The identified challenges in ML-intensive system engineering were grouped in a taxonomy which reflects the data evolution in ML pipelines.

Similarly, Hill collected insights from the development of intelligent systems by interviewing practitioners, and they reported their findings in [7]. Their outcomes present the problems faced by ML developers such as the lack of appropriate tooling concepts as well as the lack of structured procedures for their work.

Finally, Vogelsang conducted an empirical study on identifying challenges and peculiarities in the development process of ML components by interviewing data scientists [8]. The outcomes of this work points to the significant role of requirements engineering for ML component performance measurement for defining new quality requirements which fit ML aspect of a system, and for the integration of RE activities in a development process. We consider this work highly relevant for our own research which, in consequence, is directly built upon their insights and which can be seen as a continuation.

In addition to all those valuable studies, we argue that we need to explicitly consider RE in (explicit) context of holistic software engineering approaches to elaborate well on challenges and practices in RE as an integrated discipline. This consideration has, so far, not been in scope of available work. To contribute to closing this gap, we are currently conducting a case study on exploring industry experiences in the development of DCSs with the main focus on RE as an integrated discipline (considering artefacts, practices, and roles), and our vision of establishing an artefact-based RE approach for DCSs shall further contribute to effectively integrate RE.

## 3. Challenges

At the time of writing this preview, we are completing an empirical study where we interviewed practitioners form two companies followed by a questionnaire-based survey involving further 108 software engineering professionals. The scope of the study was to understand practices and challenges when handling requirements of DSCs considering their whole life-cycle.

Our study is inspired by previous work by Vogelsang et al. [8], where the authors explore challenges in RE for ML-based systems from the viewpoint of a data scientist. The idea behind our study is to extend this original work with consideration of a whole DCS development and quality assurance processes, identifying involved stakeholders, their activities, decisions they have to make, and respective artefacts they elaborate along the process. Further, we aim at finding out how the RE process is influenced by the found assets as well as challenges our respondents experience when dealing with requirements. Those insights shall foster a problem-driven research in the direction of an integrated RE for the development of DCSs.

While we cannot yet elaborate in detail on contemporary practices and challenges, we can at least enumerate the most pressing challenges we have distilled so far:

**Challenge 1:** *How can we effectively classify requirements for DCSs and how can we particularly specify quality requirements?* That is to say, what are typical classes of quality requirements and their underlying modelling concepts, and how do they differ to those important to traditional software systems?

**Challenge 2:** *What is the life cycle of (quality) requirements throughout the DCS development process?* That is, over which levels of abstraction are such requirements refined and how can they be specified at a measurable (and testable) level?

**Challenge 3:** *How do these new concepts relate to concepts already found in traditional RE (e.g., for describing functional requirements)?*

**Challenge 4:** *How can these new concepts be effectively integrated in a seamless manner?*

Note that while the first three challenges aim at laying the foundation for the engineering of requirements for DCSs, the latter aims at their process integration including relevant artefacts, roles, and activities and process elements. Only that allows for the seamless (requirements) engineering of DCSs.

## 4. Research Roadmap for an Artefact-Based RE

Motivated by the previously introduced challenges and based on our positive experiences in the use of artefact orientation, we started elaborating an artefact-based approach to RE for DCSs. A preview of that artefact model is given in Fig. 1. That model shows the relevant artefacts and their dependencies. While the thicker dashed arrows represent top-down refinement dependencies between the different content items in a traditional RE, the thinner dashed arrows depict dependencies related to the data-centric artefacts.

The goal we follow by this artefact-based philosophy is to tackle the challenges as such a model abstracts from modelling-relevant concepts and their relationships from a system-theoretical perspective and, thus, allows for seamless modelling of the RE results and their effective integration into holistic software processes. To this end, we borrow from previous work in this area and extend the Artefact Model for Domain-independent RE (short: AMDiRE), e.g.,[3]. The AMDiRE approach emerges from two decades of academia-industry collaborations. The backbone of the approach is a model of relevant artefacts which which we extent for DCSs.

Figure 1 presents a snapshot of that ongoing extension inferred from the analysis of the design techniques currently presented in the state of the art (such as in [9, 10]).

Anticipated artefacts are depicted as coloured rectangles which are associated with different levels of abstraction (i.e. layers). Originally, AMDiRE encompasses the following three layers: *Context*, *Requirements* and *System layer* [3]. We extend the traditional layers with a *Data-Centric Layer* because of our high orientation of the system behaviour on data.

We further extend the artefact model with the following artefacts driven by the analysis of industrial data-Centric component design and its integration into a system development process.

The *Context Specification* captures the operational environment of a system. New technology (e.g. AI/ML) exploration presumes new *Stakeholders* with new types of *Objectives and Goals* which affect a subset of the system characteristics. Thereby, we are also including new *Quality Attributes* such as *trustworthiness* [11]: DCSs should make a harmless decision for a human (related to safety); *explainability*[12]: the system should act in a predictable way so that humans can understand the system behaviour; *fairness*[13]: the system should make fair object distinctions and decisions where required.

The *Requirements Specification* covers the specification of the user-visible black-box behaviour and characteristics of a system. New types of non-functional requirements include *Explainability* and *Transparency Requirements* [14] which we derive from the newly defined quality attributes. Furthermore, *Fairness* and *Ethical Requirements* [14, 2] which we derive from "fairness" and "trustworthiness" quality attributes [13] and which shall support the development without any (data) discrimination. For example, DCS should make a fair and human-acceptable decision based on the processed data, e.g., the decision of a self-driving car when detecting the collision with an obstacle. However, *Fairness Requirements* and *Ethical Requirements* cover more global concerns and trade-offs, e.g., when risking harm of human beings. These requirements classes are intended to capture. Because the performance of DCSs depends on the collected data, *Data Requirements* [14] should specify all concerns about the data, e.g., definition of the data quality for the DCS under development. *Legal* and *Regulatory Requirements* [15] further specify system characteristics based on legal entities and regulations, e.g., traffic regulations for self-driving cars. Furthermore, these types of requirements aim at defining such legal aspects as data collection and storage which are handled by strict regulations for personal data protection, e.g., General Data Protection Regulation (GDPR) [16] in the European Union. Finally, *Socio-economic Requirements* cover advantages for a user along with the environmental protection [14], because DCSs bring tangible changes in human life-styles and surroundings which require proper analyses. The above-listed classes of requirements extend the definition of system quality given in the AMDiRE approach [3].

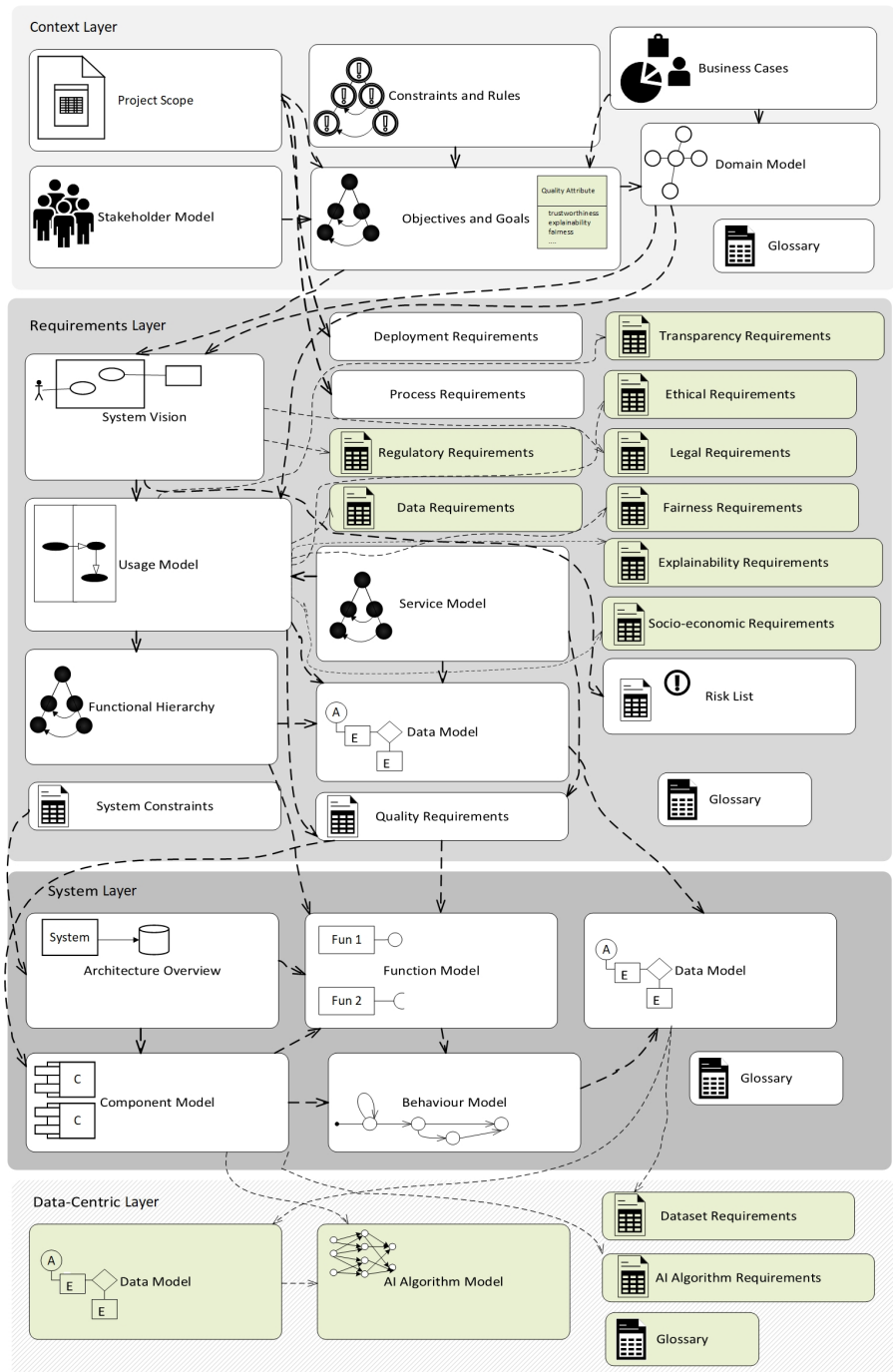

**Figure 1:** Sketch of the AMDiRE Model refined with the DCS assets and dependencies (dashed arrows)

The *System Specification*, finally, defines the solution space and considers the system in a glass box view where we decompose the system into its logical component architecture with associated functions specified in a *Functional Hierarchy* including syntactic interfaces and typed data structure (*Data Model*).

Finally, the *Data-Centric Layer* currently captures a lower-level abstraction which is decomposed into data-Centric units, e.g., AI/ML algorithms. Related artefacts include, so far, data sets used for training, testing, and verifying data-Centric algorithms; these are presented as *Data Model* and *AI Algorithm Model* in Fig. 1. At this level, requirements for data sets and for data-Centric algorithms need to be explicitly defined; namely, the specification of hyperparameters for algorithms that learn models from data derived from *Data Requirements* as well as data-Centric algorithm performance and metrics for its estimation (e.g., allowed level of object detection accuracy).

Unfortunately, we cannot discuss the different content items to the extent they deserve. Nevertheless, this snapshot, we hope, already illustrates ongoing research directions by means of artefact orientation and that it needs constant refinement and further adoption in the course of our ongoing studies to cope with the challenges experienced in practice.

## 5. Conclusion

In this paper, we framed our research direction in RE for DCSs. We presented our research preview on understanding current practices and challenges in industry and outlined our current endeavour of elaborating an artefact-based RE approach for the development of this class of systems. Our hope we associate with this research preview is to foster discussions in this direction and open the avenue for new collaborations in the RE research community.

**Acknowledgements.** This work was supported, in parts, by the KKS foundation through the S.E.R.T. Research Profile project at Blekinge Institute of Technology.